\documentclass[12pt]{iopart}
\usepackage{bm}
\usepackage{graphicx}
\begin{document}
\SUST


\title[ac losses in a finite $Z$ stack]{ac losses in a finite $Z$ stack 
using an anisotropic homogeneous-medium
approximation}

\author{John R Clem,$^1$\footnote[3]{To whom correspondence should be
addressed.} J. H. Claassen$^2$  and Yasunori Mawatari$^3$}

\address{$^1$\ Ames Laboratory and Department of Physics and Astronomy,\\
  Iowa State University, Ames, Iowa, 50011--3160, USA}

\address{$^2$\ Naval Research Laboratory, Code 6362, Washington, DC 20375, USA}

\address{$^3$\ National Institute of Advanced Industrial Science and Technology
(AIST)\\
	Tsukuba, Ibaraki 305--8568, Japan}

 \ead{clem@ameslab.gov}

\begin{abstract} 
A finite stack of thin superconducting tapes, all carrying a fixed current $I$,
can be approximated by an anisotropic superconducting bar with critical current
density
$J_c=I_c/2aD$, where $I_c$ is the critical current of each tape, $2a$ is the tape
width, and $D$ is the tape-to-tape periodicity.  The current density $J$ must
obey the constraint
$\int J dx = I/D$, where the tapes lie parallel to the $x$ axis and are stacked
along the $z$ axis.  We suppose that $J_c$ is independent of field (Bean
approximation) and look for a solution to the critical state for arbitrary
height $2b$ of the stack.  For $c<|x|<a$ we have $J=J_c$, and for $|x|<c$ the
critical state requires that $B_z=0$.  We show that this implies $\partial J/
\partial x=0$
in the central region.  Setting $c$ as a constant (independent of $z$) results
in field profiles remarkably close to the desired one ($B_z=0$ for $|x|<c$) as
long as the aspect ratio $b/a$ is not too small.  We evaluate various criteria
for choosing $c$, and we show that the calculated hysteretic losses depend only
weakly on how $c$ is chosen.  We argue that for small $D/a$ the
anisotropic homogeneous-medium approximation  gives a reasonably accurate estimate of
the ac losses in a finite $Z$ stack.  The results for a $Z$ stack can be used to
calculate the transport losses in a pancake coil wound with superconducting tape.
\end{abstract}

\submitted{\SUST}

\maketitle

\section{Introduction}

It is now possible to wind pancake coils from long lengths of
high-temperature superconducting (HTS) tape \cite{Polak06,Grilli07}. To date there
have been no theoretical calculations of the losses in this difficult geometry other
than those using a variational approach \cite{Claassen06} or via numerical
simulations \cite{Grilli07}.  A related geometry, shown in figure 1, consists of a
stack of tapes of infinite length in the $y$ direction, each carrying a total
current $I$.  This is closely related to the coil geometry but has the advantage of
computational simplicity.  The tapes of width
$2a$ in the
$x$ direction are stacked in the
$z$ direction to a height $2b$.  This problem should be
distinguished from one that has been previously considered \cite{Pardo05},
where the total current carried by all the tapes was given but the share taken
by each one could vary, as would be the case if the tapes were bundled together
to increase the current-carrying capacity of a conductor.  In a coil the current
in each winding is constrained to be the same, and we preserve this feature in
our related geometry.  We assume further that the thickness $d$ of each
superconducting film  is much smaller than the width $2a$.  This is certainly
true for 2G (second generation)  YBCO tape \cite{Goyal05}, where the
tape-to-tape periodicity $D$ is governed by the thicknesses of the insulator,
substrate, and buffer layers.

\begin{figure}
\centering
\includegraphics[width=8cm]{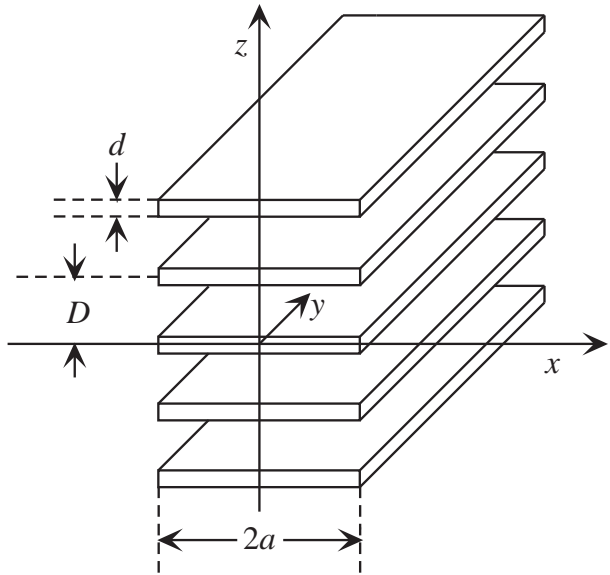}
\caption{Finite $Z$ stack: a stack of superconducting strips of infinite length
in the
$y$ direction, each carrying  current $I$.  The overall height of the stack is
$2b$.  }
\label{Fig01}
\end{figure} 

Both Mawatari \cite{Mawatari97} and 
M\"uller \cite{Muller97,Muller99}\footnote{Reference
\cite{Muller99} also
corrects typographical errors in expressions for the ac losses given in 
\cite{Muller97}.}  considered this problem in the limit $b \to \infty$
 and obtained analytic
expressions for the fields and currents.  Both authors noted that in the limit $D
\ll a$  the solutions approach those for a uniform infinite slab of width $2a$
carrying an average current density 
$I/2aD$.  In other words, the stack becomes equivalent to a homogenous
superconducting slab with critical current density $J_c=I_c/2aD$, where $I_c$ is
the critical current in  each
tape.  We expect that in practical applications the ratio $D/a$ will lie in the
range 0.01-0.2.  In figure 2 we show exact calculations of the ac losses as in 
\cite{Mawatari97}, normalized to the ac losses calculated using the 
homogeneous approximation, as a function of
$D/a$.  It can be seen that the homogeneous approximation is reasonably accurate
for small $D/a$ and large $I/I_c$.  Specifically, if we restrict ourselves to
currents of amplitude greater than 0.2 $I_c$, this approximation gives better
than  20\% accuracy if $D/a < 0.2$.  From an engineering perspective
this sort of accuracy is usually
adequate, especially since the error is in the right direction (overestimating,
rather than underestimating, the dissipation).

\begin{figure}
\centering
\includegraphics[width=8cm]{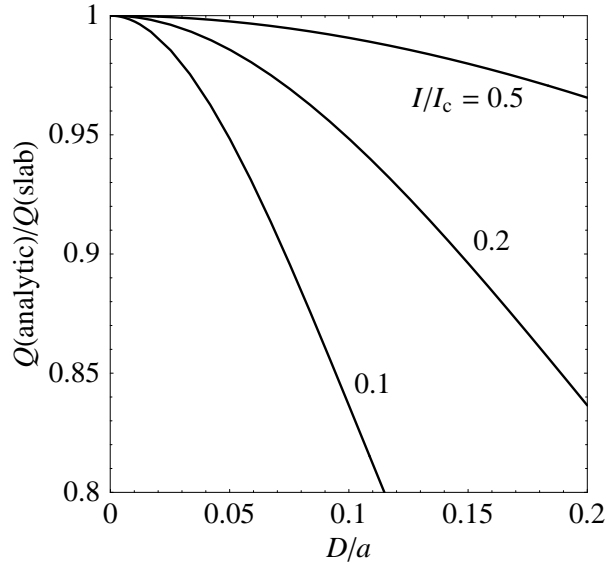}
\caption{ac losses in an infinite $Z$ stack, calculated from the analytic
solutions in \cite{Mawatari97}, normalized to
the losses in the equivalent uniform slab, at various current amplitudes $I$ and
stack periodicities $D$.}
\label{Fig02}
\end{figure} 

At present there are no analytic solutions available for the problem of a finite
stack of conductors.  To initially approach this problem it makes sense to use
an approach that has some of the features of a homogeneous model.  However,
our model must also account, at least approximately, for the  screening by
subcritical portions of the superconducting strips.  It is  likely that the
error in this approach will be similar to that of the infinite stack; see figure
2.  The current density
$J_y$ and magnetic induction
$\mathbf B$ are averaged over a volume $D^3$; that is, we use only macroscopic
values of these quantities.  To model the constraint of constant total current
in each tape, we require that $\int J_y dx=I/D$ for all $|z|<b$.
In section 2 we use this anisotropic homogeneous-medium approximation to
calculate the ac losses of a finite $Z$ stack of superconducting tapes.  We
discuss and summarize our results in section 3.

\section{Anisotropic homogeneous-medium approximation} 

We consider a sample initially in the virgin (magnetic-flux-free) state and
examine the initial penetration of magnetic flux as current is applied in the $y$
direction.  We  anticipate that, similar to case of an infinite slab, we will
have a region
$c<|x|<a$ with
$J_y=J_c$.  For simplicity, we use the  Bean \cite{Bean62,Bean64} critical state
model, in which
$J_c$ is independent of field.  Unlike the behavior in a homogeneous infinite
slab, however, in principle we should  allow for $c$ to vary as a function of
$z$. 
Further, we cannot assume that $J_y=0$ and $\mathbf B=0$ in the region
$|x|<c(z)$, as is the case for the homogeneous infinite slab.
It is known from studies of the critical state model in an isolated
superconducting strip \cite{Brandt93b,Zeldov94b} that no significant amount of
magnetic flux can penetrate subcritical portions of the strip (i.e., $B_z=0$
wherever $J_y<J_c$); this is also true for each of the strips in the
$Z$ stack. On the other hand, a finite
$B_x$  is allowed, since magnetic flux can thread between the superconducting
layers from the ends of the tapes without fully penetrating any superconductor. 
This leads to important constraints on $\mathbf B$ and $J_y = J_m$ in the middle
region
$|x|<c(z)$: Since $\nabla \cdot \mathbf B=0$, we must have $ \partial
B_x/\partial x=0$, such that  $B_x$ depends only on $z$.  Ampere's law requires
that 
$\mu_0 J_m=\partial B_x/\partial z-\partial B_z/\partial x$.  Since the second
term on the right-hand side is zero and the first depends only on $z$, we
conclude that $J_m$ can depend only on $z$.  Thus the
current density $J_y$ as a function of $x$ has a step-function character, with
the values $J_m$ for $|x|<c(z)$ and $J_c$ for $|x| > c(z)$.  To have a fixed
total current in each layer we require
\begin{equation}
J_m/J_c=1-(a/c)(1-I/I_c).
\label{Jm}
\end{equation}
For finite values of $b$, the current density $J_c$
in the region $c < x < a$ contributes, via the Biot-Savart law, a positive value
of $B_z(c,0)$, while the current density $J_c$ in the region $-a < x < -c$
contributes a negative value
 of smaller magnitude.  In order to make $B_z(c,0) = 0$, the current
density in the region $-c < x < c$ must obey $J_m > 0$, so that it makes a
negative contribution to $B_z(c,0)$, thereby cancelling the net positive
contribution from the currents in the regions for which $c < |x| < a$.
Since $0 < J_m < J_c$ and $0 < c < a,$
 we thus see that
$c/a$ can vary in the range from $(1-I/I_c)$ to 1. 
In the limit as $b \to \infty$, we must find that $J_y/J_c \to 0$ for   $|x|<c$
and that
$c/a \to 1-I/I_c$. The theoretical problem thus reduces to finding a  $c(z)$
that  yields macroscopic fields consistent with the above requirements of the
critical state.  This means that we must have a region defined by $|x|<c(z)$
where
$B_z=0$.  

Our primary goal in this paper is to calculate the
hysteretic ac losses in a $Z$ stack.  
Using the above approach, once we obtain the solutions for $B_z(x,z)$, we begin
by finding $Q_{init}'$, the energy per unit length dissipated upon initial
penetration of magnetic flux, i.e., when the  current in each tape is raised 
from zero to a maximum value
$I < I_c$, starting from the virgin state (no trapped magnetic fields in the
superconductor).  To derive $Q_{init}'$, we (a) integrate $\mathbf J \cdot
\mathbf E$ over the cross section of the stack, (b) neglect  the relatively
small losses in the tapes in the middle region, $|x| < c(z)$, where $J_y < J_c$,
$B_z(x,z) = 0$, and $\mathbf E = 0$, (c) note that $J_y = J_c$ in the outer
regions,
$c(z) < |x| < a$, (d) apply Faraday's law, $\nabla \times \mathbf E = -\partial
\mathbf B/\partial t$, (e) integrate over time as the
magnetic induction $\mathbf B$ increases from zero and reaches its final value,
(f) make use of the symmetry that the losses are the same in all four quadrants
of the
$xz$ plane, and (g) do a partial integration over
$x$.  The result is \cite{Claassen06,Muller97,Mawatari06,Mawatari07}  
\begin{equation}
Q_{init}' = -4 J_c \int_0^b dz \int_{c(z)}^a dx (a-x) B_z(x,z).
\label{Qinit}
\end{equation}
The physical interpretation of this formula is that $Q_{init}'$ is the summation
of the energy dissipated by vortices as they move a distance $a-x$ 
from the edge to their final positions; the force per unit length is $\phi_0
J_c$, where
$\phi_0 = h/2e$ is the superconducting flux quantum, and the density of vortices 
is $B_z(x,z)/\phi_0$.  The current and field distributions during the initial
penetration of magnetic flux are not the same as those that occur during one quarter
of the ac cycle.  Nevertheless, it can be shown \cite{Norris70,Halse70} that $Q',$
the hysteretic ac loss per cycle per unit length, is given by  $Q' = 4Q_{init}'$.  

For an infinite slab of thickness $2a$, the magnetic induction upon initial
penetration for $x>c =a(1-I/I_c)$ is $B_z(x) = -\mu_0 J_c (x-c)$, and the
hysteretic ac loss per unit length associated with a cross-sectional area $4ab$
is 
\begin{equation}
Q_{inf}' = \frac{8}{3} \mu_0 J_c^2 a^3 b (I/I_c)^3.
\label{Qinf}
\end{equation}
 
A good starting point for the calculation of the magnetic induction inside
the $Z$ stack is the assumption that
$c={\rm const}$, independent of $z$, and this is the approximation that we shall
use for the remainder of this paper.
Expressions for the magnetic induction $\mathbf B(x,z) = \hat x B_x(x,z) + \hat z
B_z(x,z) =
\nabla \times \mathbf A$ and the corresponding vector potential $\mathbf
A = \hat y A_y(x,z)$ generated by a current density in the stack
$J_y=J_m =[1-(a/c)(1-I/I_c)]J_c$ for
$|x|<c$ and  $J_y = J_c$ for $c < |x| < a$ can be
obtained by using the Biot-Savart law and integrating over the cross section of
the stack, $|x| < a$ and  $|z| < b$.  Results obtained for a constant value
of $c$ are given in Appendix A and Appendix B, and we have evaluated them
numerically using Mathematica \cite{Math}. It can be shown that these expressions
can never exactly satisfy the requirement that $B_z(x,z) = 0$ for all $|x| > c$
and
$|z| < b$. However, if we choose $c/a$ to make the average of $B_z(c,z)$ over the
region $0 < x < c$ and
$0 < z < b$ equal to zero [see  (\ref{Aintegral})], we find that 
$B_z(x,z)
\approx 0$  for all
$|x| < c$ and
$|z| < b$ with an accuracy that improves as $b \to \infty$.
Figure 3 shows the dependence of $c$ upon the current $I$ for
various stack heights $2b$, and figure 4 shows 
corresponding plots of $J_m$ (\ref{Jm}) vs
current.  As expected,  in
the limit as
$b/a
\to
\infty$, $c$ approaches the limiting value
$a(1-I/I_c)$ for an infinite slab of thickness $2a$ and $J_m$ approaches zero. 

\begin{figure}
\centering
\includegraphics[width=8cm]{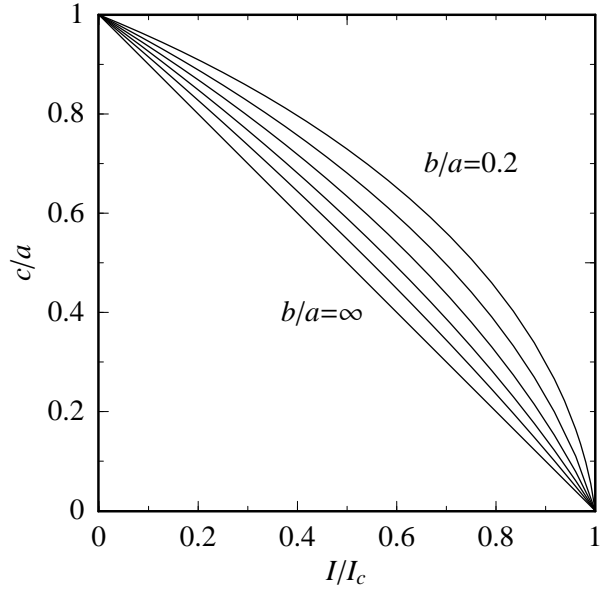}
\caption{The constant $c$ (in units of $a$) obtained from  (\ref{Aintegral})
as a function of
$I/I_c$ for stack aspect ratios $b/a = 0.2, 0.5, 1, 2, 5,$ and $\infty$ (top to
bottom). For the latter case, $c/a = 1-I/I_c$. }
\label{Fig03}
\end{figure} 

\begin{figure}
\centering
\includegraphics[width=8cm]{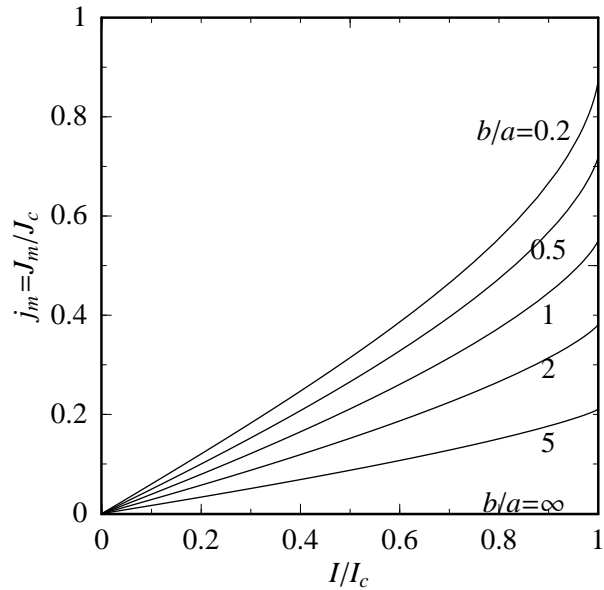}
\caption{The reduced current density $j_m=J_m/J_c$ in the middle region $|x| < c$
and
$|z| < b$ as a function of $I/I_c$, calculated from (\ref{jm}) and
(\ref{Aintegral}), where
$c$ is assumed to be independent of
$z$, for stack aspect ratios $b/a = 0.2, 0.5, 1, 2, 5,$ and $\infty$.  For
the latter case, $j_m = 0$.}
\label{Fig04}
\end{figure} 

\begin{figure}
\centering
\includegraphics[width=8cm]{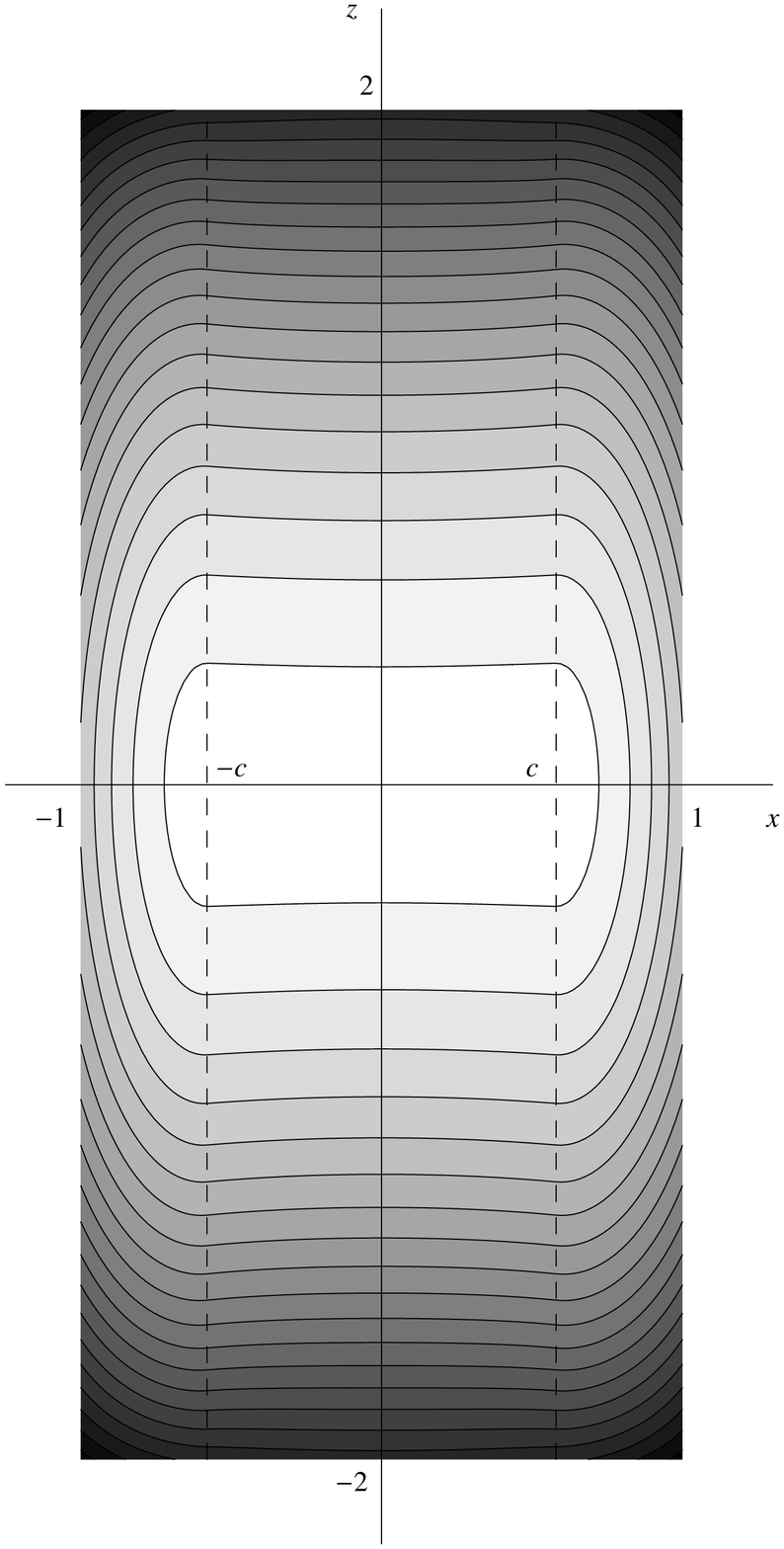}
\caption{Contour plot of $A_y(x,z)$ vs $x$ and $z$ (in units of $a$) inside the
$Z$ stack for
$I/I_c = 0.5$ and
$b/a = 2$.  The contours
correspond to magnetic field lines flowing in the clockwise direction.
 The vertical
dashed lines mark the boundaries of the middle region at $|x|/a = c/a = 0.5895$.}
\label{b20}
\end{figure}

\begin{figure}
\centering
\includegraphics[width=8cm]{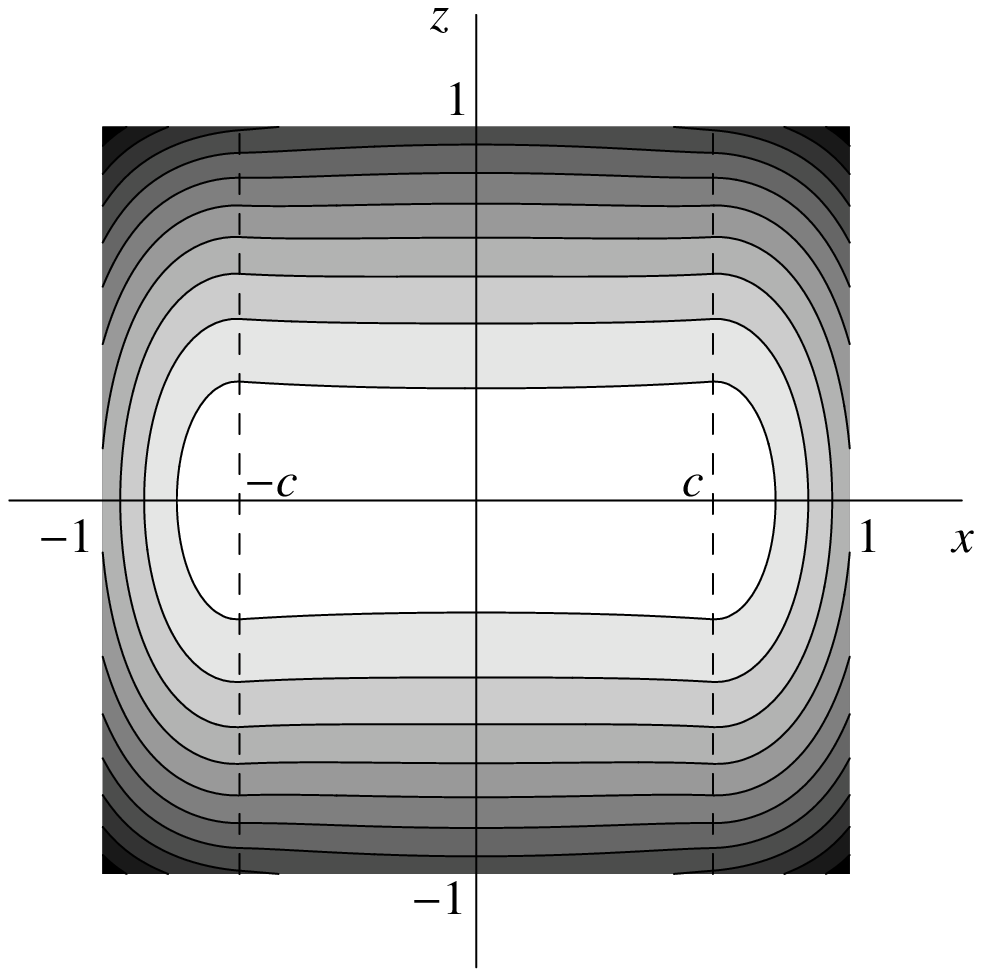}
\caption{Contour plot of $A_y(x,z)$ vs  $x$ and $z$ (in units of $a$) inside the
$Z$ stack for
$I/I_c = 0.5$ and
$b/a = 1$.  The contours
correspond to magnetic field lines flowing in the clockwise direction.
 The vertical
dashed lines mark the boundaries of the middle region at $|x|/a = c/a = 0.6338$.}
\label{b10}
\end{figure}

\begin{figure}
\centering
\includegraphics[width=8cm]{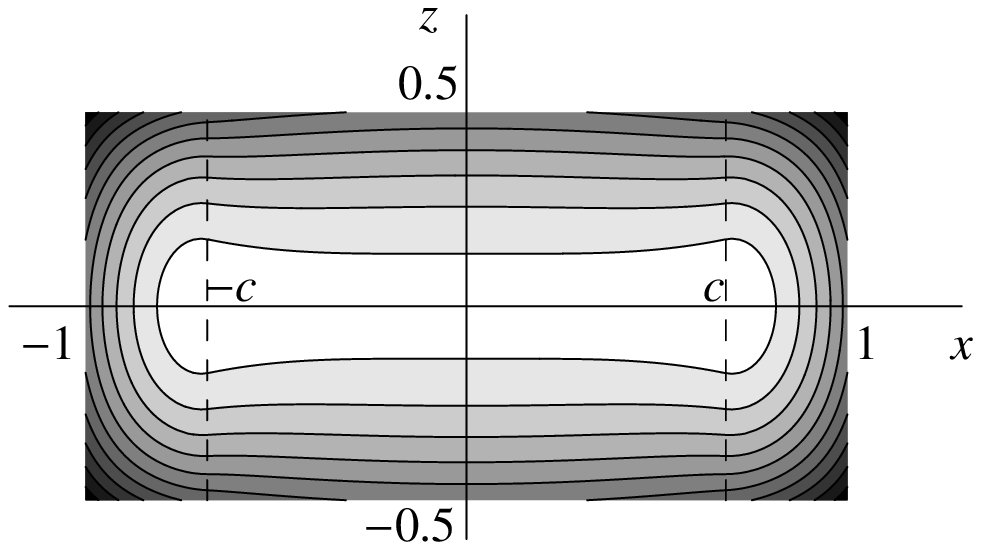}
\caption{Contour plot of $A_y(x,z)$ vs $x$ and $z$ (in units of $a$) inside the
$Z$ stack for
$I/I_c = 0.5$ and
$b/a = 0.5$.  The contours
correspond to magnetic field lines flowing in the clockwise direction.
 The vertical
dashed lines mark the boundaries of the middle region at $|x|/a = c/a = 0.6809$.}
\label{b05}
\end{figure}

\begin{figure}
\centering
\includegraphics[width=8cm]{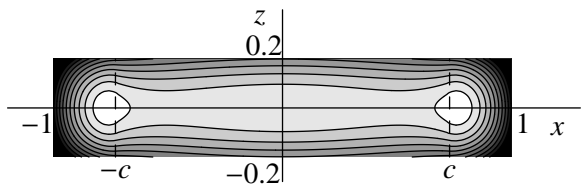}
\caption{Contour plot of $A_y(x,z)$ vs $x$ and $z$ (in units of $a$) inside the
$Z$ stack for
$I/I_c = 0.5$ and
$b/a = 0.2$.  The contours
correspond to magnetic field lines flowing in the clockwise direction.
 The vertical
dashed lines mark the boundaries of the middle region at $|x|/a = c/a = 0.7292$.}
\label{b02}
\end{figure}

Figures \ref{b20}, \ref{b10}, \ref{b05}, and \ref{b02} show contours of constant
$A_y(x,a)$  for values of $c$ obtained from 
(\ref{Aintegral}).  Since these contours correspond to magnetic field lines, an
exact solution would have  all these contours
parallel to the $x$ axis in the middle region $|x| < c$, where we should have
$B_y(x,z) = 0$.  The degree to which these contours meet this criterion is one
measure of the accurary of our method of approximation.  Despite the simplicity
of our approximation, the magnetic field lines are remarkably straight in the
middle region  $|x| < c$ between the vertical dashed lines, especially for stack
aspect ratios $b/a \ge 1$. However, as can be seen in Figs.\ 7 and 8, the
field lines deviate from the desired straightness in the middle region for
smaller aspect ratios.

In  figure 9 we plot $B_z(x,z)$ at various heights $z$
above the center line for $b=a$ and $I/I_c$ = 0.5.  Note that although
$|B_z(x,z)|$ is generally much smaller in the middle region $|x| < c$ than in
the regions carrying a critical current ($c < |x| < a$),  it is not precisely
equal to zero, as would be the case for an exact solution.  Note also that
although  $|B_z(x,z)|$ is very nearly zero in the middle region $|x| < c$ for $z
= 0$, it deviates from this behavior as we move away from the center line. 

\begin{figure}
\centering
\includegraphics[width=8cm]{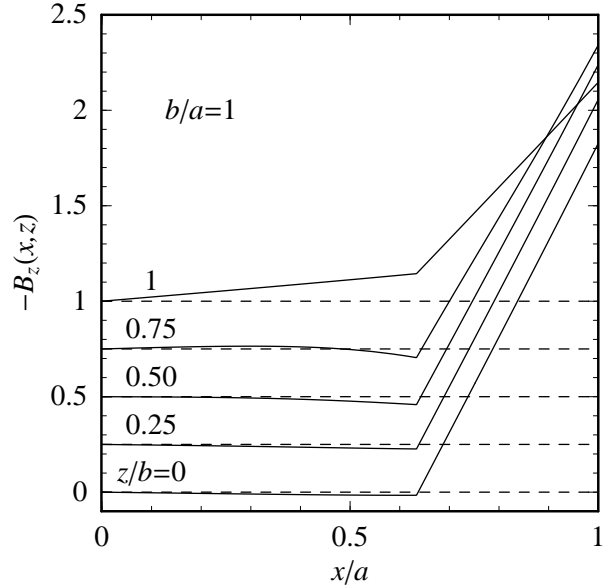}
\caption{$-B_z(x,z)$  in units of $\mu_0 a J_c / 2\pi$, calculated using a
constant value of $c$ at five heights in the stack, for a square stack ($b/a =
1$) with $I/I_c = 0.5$ and $c/a =0.6338 a$ from  (\ref{Aintegral}). The
curves are offset vertically; by symmetry, 
$B_z(0,z)=0$ for any $z$, as marked by the horizontal dashed lines.   }
\label{Fig09}
\end{figure}

Several different criteria could have been used to determine the constant $c$. 
Our  choice, based on the best appearance of the contours of constant $A_y(x,z)$,
is to use the procedure given in Appendix B and  (\ref{Aintegral}).  Let us
call this criterion (i).  We also could have determined $c$ by choosing
$B_z(x,z)$ or $B_{zx}(x,z) = \partial
B_z(x,z)/\partial x$ to be zero at various locations.  For example, listed in
Table \ref{1} are  values of $c/a$ determined using the following criteria: (i) 
 (\ref{Aintegral}), (ii) $B_z(c,0)=0$, (iii) $B_{zx}(0,0) = 0$,   (iv)
$B_{zx}(0,b/2) = 0$, and (v) $B_{zx}(0,b) = 0$.
\begin{table}
\centering
\caption{\label{1}The constant $c$ (in units of $a$) for $I/I_c =
0.5$ determined by the five different criteria discussed in the text.}
\vspace{1 mm}
\begin{tabular}{cccccc}
\hline
{$b/a$} & {i} & {ii} & {iii} & {iv} & {v} \\ 
\hline
$\infty$ & 0.5000 & 0.5000 & 0.5000 & 0.5000 & 0.5000\\
10 & 0.5286 & 0.5163 & 0.5163 & 0.5218 & 0.5081\\
5 & 0.5478 & 0.5331 & 0.5332 & 0.5431 & 0.5163\\
2 & 0.5895 & 0.5803 & 0.5816 & 0.5927 & 0.5416\\
1 & 0.6338 & 0.6372 & 0.6383 & 0.6351 & 0.5816\\
0.5 & 0.6809 & 0.6996 & 0.6821 & 0.6727 & 0.6383\\
0.2 & 0.7292 & 0.7765 & 0.7025 & 0.6995 & 0.6902\\
\hline
\end{tabular}
\end{table}
Table \ref{2} gives values of the ratio
\begin{equation}
\int_0^c dx \int_0^b dz B_z(x,z)/cbB_z(a,0)
\end{equation}
for $I/I_c=0.5$ and various values of $b/a$ using the five different criteria. 
\begin{table}
\centering
\caption{\label{2}The average of $B_z(x,z)$ over the quadrant $0 < x
< c$ to 
$0 < z < b$  divided by $B_z(a,0)$ for
$I/I_c = 0.5$ determined by the five different criteria discussed in the text.}
\vspace{1 mm}
\begin{tabular}{cccccc}
\hline
{$b/a$} & {i} & {ii} & {iii} & {iv} & {v} \\ 
\hline
$\infty$ & 0.0000 & 0.0000 & 0.0000 & 0.0000 & 0.0000\\
10 & 0.0000 & -0.0122 & -0.0122 & -0.0068 & -0.0204\\
5 & 0.0000 & -0.0149 & -0.0148 & -0.0047 & -0.0318\\
2 & 0.0000 & -0.0098 & -0.0084 & 0.0034 & -0.0511\\
1 & 0.0000 & 0.0039 & 0.0051 & 0.0015 & -0.0588\\
0.5 & 0.0000 & 0.0215 & 0.0014 & -0.0094 & -0.0486\\
0.2 & 0.0000 & 0.0499 & -0.0275 & -0.0306 & -0.0401\\
\hline
\end{tabular}
\end{table}
To show the sensitivity of
$B_z(x,0)$ to the choice of the constant
$c$, we show in figure 10 plots of $-B_z(x,0)$ and $-B_z(x,b)$ vs $x/a$ for $b=a$,
$I/I_c$ = 0.5, and  values of $c$ determined using three different criteria,
(i), (ii) and (v).  With criteria (i) and (ii), $B_z(x,z)$ is very nearly
equal to zero for $|x|<c$ and  $z = 0$ but deviates more strongly at $z = b$,
while the opposite is true for criterion (v).   This provides
additional evidence that the approximation of choosing
$c(z)$ to be independent of $z$ can never yield $B_z(x,z) = 0$ throughout the
middle region, $|x| < c$ and $|z| < b$.  Note, however, that for $c < |x| < a$,
$B_z(x,z)$ is very nearly the same for all three criteria.

\begin{figure}
\centering
\includegraphics[width=8cm]{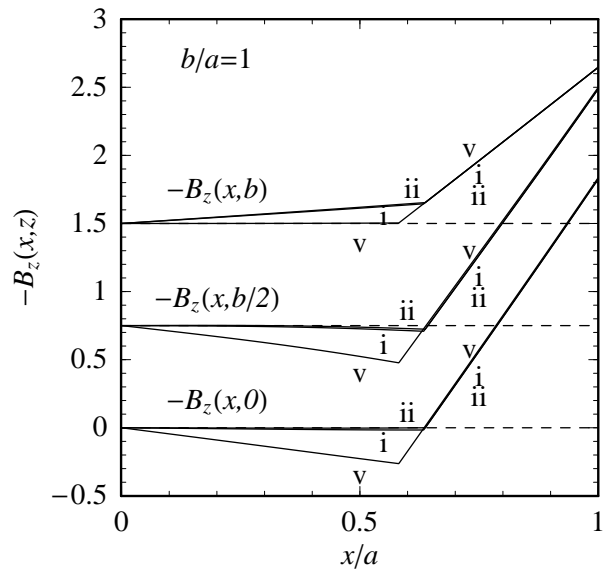}
\caption{$-B_z(x,0)$, $-B_z(x,b/2)$, and $-B_z(x,b)$
 in units of
$\mu_0 a J_c / 2\pi$ for a square stack ($b/a = 1$) with $I/I_c = 0.5$,
calculated using three different criteria (see text) to determine $c/a$:
(i) 0.6338, (ii) 0.6372, and (v) 0.5816. For $x/a <0.5816$, criterion ii yields
the top curve, criterion i the middle curve, and criterion v the bottom curve.
For
$x/a > 0.6372,$ the order is reversed, but all three criteria give very nearly
the same value of
$B_z(x,z)$.  The curves are offset vertically; by
symmetry, 
$B_z(0,z)=0$ for any $z$, as marked by the horizontal dashed lines.   }
\label{Fig10}
\end{figure}

For small $b/a$, the $c$ = constant approximation is less successful in
approximating the true fields in a $Z$ stack.  Figure 11 shows $-B_z(x,0)$ along
the center line ($z = 0$) for various currents with $b/a$ = 0.5, and figure 12
shows the same for $b/a$ = 0.2.  In the latter case the fields in the
middle portion of the stack differ significantly from our desired condition $B_z
= 0$.  Moreover, as can be seen in figure 12 for $b/a = 0.2$ and $I/I_c = 0.2$,
calculations using criterion (i) yield values of
$-B_z(x,0) < 0$ for some values of
$x$ in the penetrated region $c < x < a$. Since these negative values are weighted by
the factor $(a-x)$ in  (\ref{Qinit}), the losses calculated in the limit as $I/I_c
\to 0$ using criterion (i) even become negative for $b' = b/a < 0.0457$, an
unphysical result.  On the other hand, we expect the errors in the losses due to such
negative values of
$-B_z(x,0)$ to be small for practical values of $I/I_c > 0.2$ when $b' = b/a
\sim 1$.

\begin{figure}
\centering
\includegraphics[width=8cm]{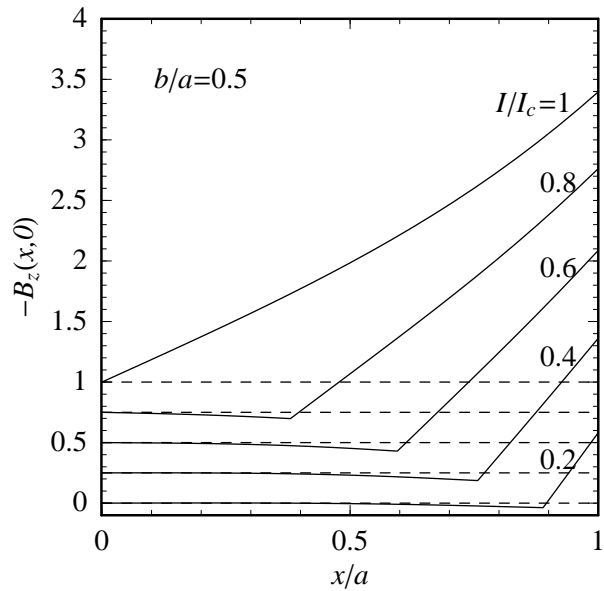}
\caption{$-B_z(x,0)$  in units of $\mu_0 a J_c / 2\pi$ vs $x/a$ for various
currents in a stack of moderately low aspect ratio, $b/a=0.5$. The values of
$c/a$ obtained from  (\ref{Aintegral}) are, for $I/I_c =$ 0.2, 0.8889; 0.4,
0.7572; 0.6, 0.5952; 0.8, 0.3798; and 1, 0. (See figure 3.)}
\label{Fig11}
\end{figure}
 
\begin{figure}
\centering
\includegraphics[width=8cm]{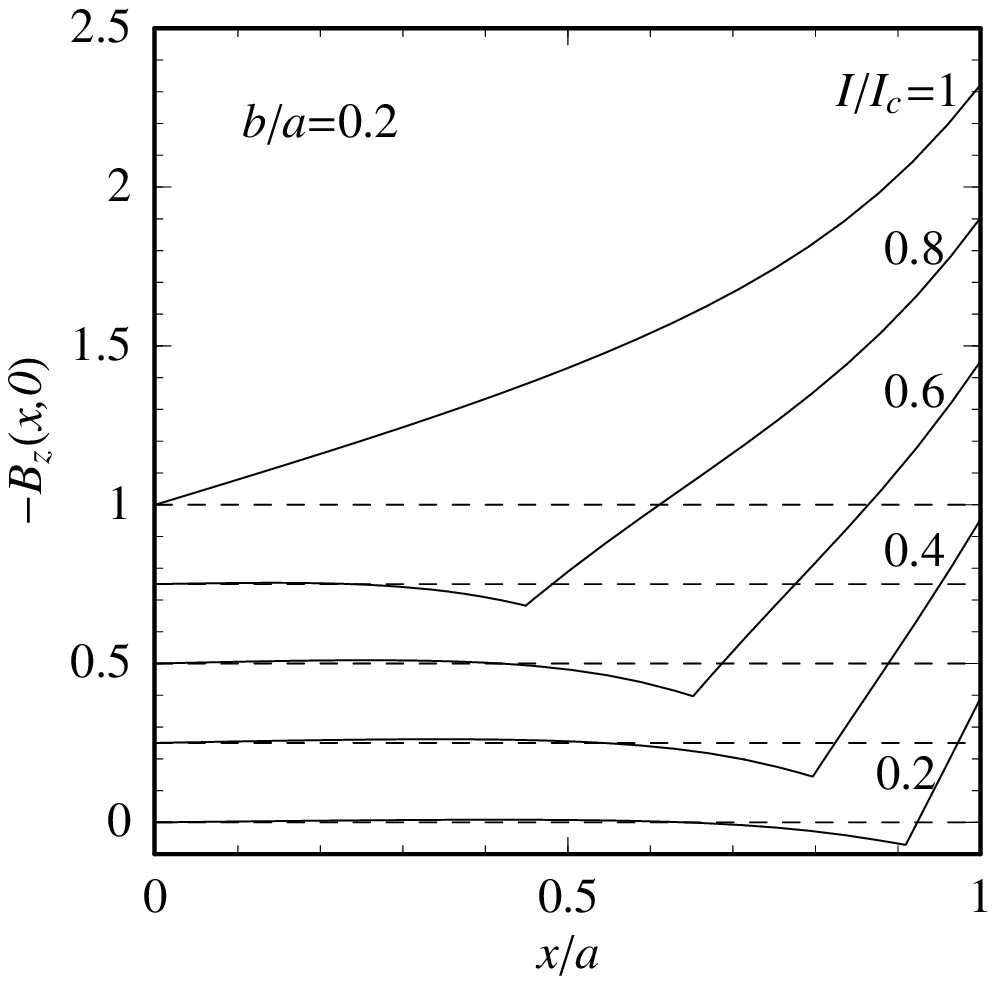}
\caption{$-B_z(x,0)$ 
in units of $\mu_0 a J_c / 2\pi$ vs $x/a$ for  various currents in a stack of
low aspect ratio, $b/a=0.2$.  The values of
$c/a$ obtained from  (\ref{Aintegral}) are, for $I/I_c =$ 0.2, 0.9094;
0.4, 0.7967; 0.6, 0.6517; 0.8, 0.4489; and 1, 0. (See figure 3.)}
\label{Fig12}
\end{figure}

Figure 13 shows calculated values of $-B_z(x,0)$ vs $x/a$  for various aspect
ratios $b/a$.  Note that for decreasing values of $b/a$, the profiles of 
$-B_z(x,0)$ become shallower and that the critical region  where $J_y = J_c$ is
closer to the edges at $x = a$.  It is clear from inspection of 
(\ref{Qinit}) that this behavior  will lead to lower losses in $Z$ stacks with
smaller aspect ratios. 
We have used the above method for obtaining $B_z(x,z)$ in 
(\ref{Qinit}) to determine the hysteretic ac loss per cycle per unit length.

\begin{figure}
\centering
\includegraphics[width=8cm]{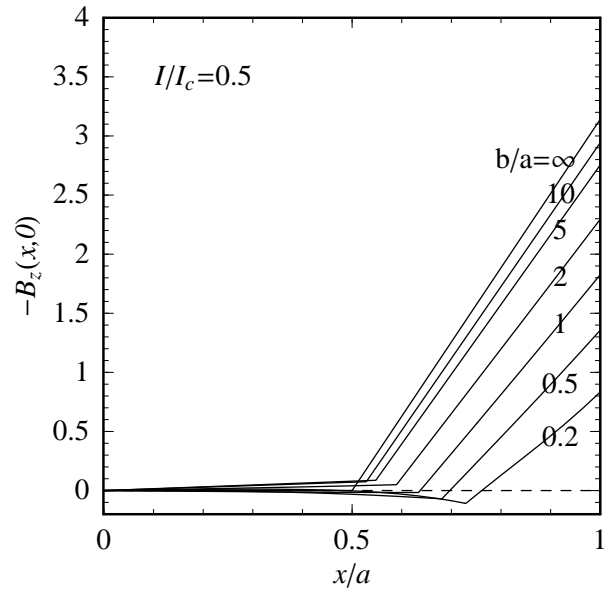}
\caption{$-B_z(x,0)$ 
in units of $\mu_0 a J_c / 2\pi$ vs $x/a$ for $I/I_c = 0.5$ and various aspect
ratios, $b/a=0.2$, 0.5, 1, 2, 5, 10, and $\infty$ (bottom to top). }
\label{Fig13}
\end{figure}

\begin{figure}
\centering
\includegraphics[width=8cm]{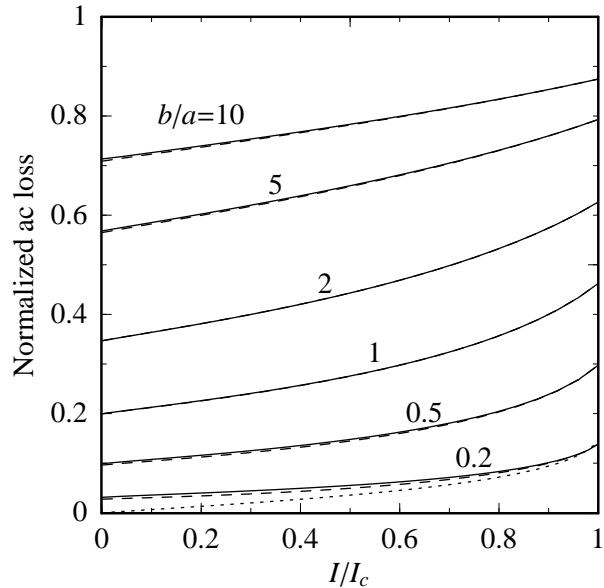}
\caption{$Q'$, the ac loss per cycle per unit length in a $Z$ stack in the
anisotropic homogeneous-medium approximation, normalized to $Q_{inf}'$ (\ref{Qinf}),
the ac loss per cycle per unit length for an equivalent cross section of an
infinite slab, using criterion (i) solid and (ii) dashed. 
The dotted curve shows the results of Norris \cite{Norris70} 
(\ref{QNorris}) for an isolated thin strip of thickness
$2b = 0.4 a$.}
\label{acloss}
\end{figure}

Shown in figure \ref{acloss} are our results for $Q'$, the loss per cycle per unit
length, normalized to $Q_{inf}'$ (\ref{Qinf}),
the loss per cycle per unit length of an infinite slab within a cross section
$4ab$.  
Note  that as the aspect ratio $b/a$ increases, the ac loss
per cycle converges slowly toward that of an infinite slab, but it can be
significantly lower when $b/a \sim 1.$  Included in figure \ref{acloss} are our
results for
$b/a=0.2,$  although our approach becomes increasingly
problematic at low aspect ratio,  as discussed above. 
The solid and dashed curves show the dependence of
the calculated loss on the criterion used to choose $c/a$.  Clearly there is
very little difference, which can be understood by inspecting the curves for
$-B_z$ vs $x$ in figure 10.  The behavior in the region $|x| > c$, which
enters the loss calculation, is only weakly influenced by the choice of $c$.
 
For comparison, the dotted curve in figure \ref{acloss} shows the 
Norris \cite{Norris70} result for
$Q_{strip}'$, the hysteretic loss per cycle per unit length for an isolated
thin, flat  strip of thickness
$2b$, normalized to $Q_{inf}'$  (\ref{Qinf}), where
\begin{equation}
Q_{strip}' = \frac{ 16 \mu_0 J_c^2 a^2 b^2}{\pi}[(1-F)\ln(1-F) +(1+F)\ln(1+F)-F^2]
\label{QNorris}
\end{equation}
and $F = I/I_c$.
For $F \ll 1$, 
\begin{equation}
Q_{strip}'/ Q_{inf}' \approx \Big(\frac{b}{\pi a}\Big) \frac{I}{I_c}.
\label{QNorrisRatio}
\end{equation}

The intercepts in figure 14 of the normalized ac loss in the limits $I/I_c \to 1$
and 
$I/I_c \to 0$ are plotted vs $b/a$ as the upper and lower curves in figure 15. 
The details of how to calculate
$Q'$ in these two limits are given in Appendixes C and D.  When $I/I_c \to 1$,
$Q'$ is independent of the criterion used to determine the parameter $c$, and
when $I/I_c \to
0$, $Q'$ depends only very weakly upon the criterion; in figure 15, the results
calculated using criteria (i) and (ii) are almost indistinguishable.

\begin{figure}
\centering
\includegraphics[width=8cm]{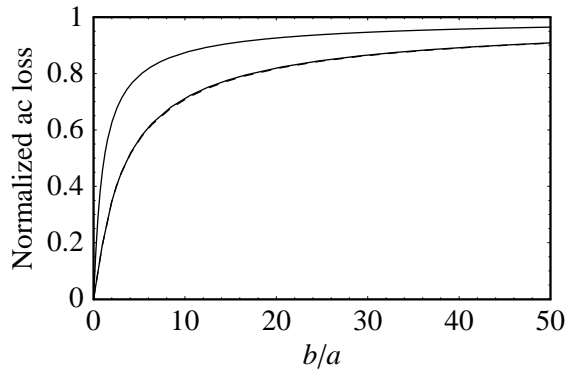}
\caption{$Q'$, the ac loss per cycle per unit length in a $Z$ stack in
the anisotropic homogeneous-medium approximation, normalized
to
$Q_{inf}'$ (\ref{Qinf}), the ac loss per cycle per unit length for an
equivalent cross section  of an
infinite slab, vs $b/a$.  The upper curve shows the result $R_1$ (\ref{R1}) in
the limit  in the  $I/I_c \to 1,$ and the lower curves show the result $R_0$
(\ref{R0}) in the limit
$I/I_c \to 0$ using criteria (i) solid and (ii) dashed.}
\label{aclosslimits}
\end{figure}

In the above calculations we have neglected the losses due to the currents and parallel
ac fields ($B_x$) in the middle region $|x| < c$. In Appendix E we provide
equations that can be used to estimate the middle-region losses.  For typical tape
dimensions and $I/I_c > 0.2$ we find that the middle-region losses are
several orders of magnitude smaller than the losses at the edges $(c < |x|
< a)$.

\section{Discussion and Summary}

Making use of what we have
called the  anisotropic homogeneous-medium approximation,  we have
introduced in this paper a theoretical framework for estimating the ac
losses in a finite $Z$ stack of superconducting tapes via straightforward analytic
calculations.  Our results yield $Q'$, the hysteretic loss per cycle per unit length
of a $Z$ stack of total height $2b$, where the tapes have width $2a$.  We have
found it useful to compare our results with $Q_{inf}'$, the hysteretic ac loss per
cycle per unit length for an equivalent cross section of an infinite slab.  Our
calculation is intended for application to the problem of calculating the hysteretic
ac losses of a pancake coil, where the stack of tapes does not extend to infinity but
rather curves back on itself.  As long as the radius $R$ of the coil is much greater
than
$a$ or
$b,$ the field solutions should not differ significantly from what we have calculated,
and the total hysteretic loss per cycle should be well approximated by $2 \pi R Q'$.

For a stack of many tapes, $Q'$ is much larger than that for a single
tape \cite{Grilli07}, and this can be understood most simply by noting that  $Q' =
4 Q_{init}'$, where $Q_{init}'$ is given in  (\ref{Qinit}).  Not only is the
magnitude of the magnetic flux density $B_z(x,z)$, which appears on the right-hand
side of this equation, much larger than that generated by a single tape because of
the superposition of the field contributions from all the tapes, but also 
$B_z(x,z)$ is integrated in the $z$ direction over a much greater height.  For a
stack of aspect ratio $b/a \sim 1$ it is more appropriate to compare $Q'$ with
$Q_{inf}'$, the  hysteretic ac loss per
cycle for an equivalent cross section of an infinite slab.  For all finite values of
$b/a$ we find $Q' <  Q_{inf}'$.  Referring to  (\ref{Qinit}) and the field
lines in Figs.\ 5-8, we see that the reason for this is that the field 
bends around the corners of the
finite $Z$ stack, and therefore $ |B_z(x,z)|$ is always less than
the corresponding quantity
$|B_z(x)|$ for the infinite slab.

Grilli and Ashworth \cite{Grilli07} have recently presented loss data for a multiturn
pancake coil, which should be amenable to analysis using the formalism presented here. 
To calculate the losses, they used a finite-element method, which apparently
requires considerable computational resources.   While we believe that our
analytic approach has the advantage of efficiently yielding a loss prediction of
sufficient accuracy for many applications, it would be of interest to compare results
obtained using these two different approaches.  One advantage of detailed
finite-element calculations such as those discussed in \cite{Grilli07} is
the capability  of self-consistently incorporating the $B$ dependence of the critical
current density $J_c(B)$.  Our approach has made use of the assumption that $J_c$ is a
constant, independent of $B$.  However, it would be possible for us to account crudely
for the
$B$ dependence of $J_c(B)$ by using a model for this dependence and replacing the
constant
$J_c$ at each current amplitude $I$ with $J_c(B_m)$, where $B_m$ is the maximum
magnitude of
$B_z(x,z)$ at
$(x,z) = (a,0)$.

In section 2 and Appendix D we have pointed out that our anisotropic
homogeneous-medium approximation is not accurate for small values of $b/a$, i.e., for a
small number of superconducting layers in the stack.  One reason for this is that we
have assumed  that each layer carries a constant average current density
$J_m$ in the middle region $|x| < c$.  The motivation for this assumption is that $B_z$
must be zero there and $B_x$ between each pair of superconducting layers must be
independent of $x$.  However, the magnetic induction $B_x$ at the top and bottom
surfaces of the
$Z$ stack is not subject to this constraint but in general depends upon $x$,
and consequently the screening sheet currents on the top surface of the top tape and
bottom surface of the bottom tape also depend upon $x$. It seems likely that our
failure to account for these sheet currents, which become relatively more important
for a small number of layers, is the major reason for the problems with the present
theory for small values of $b/a$.

\ack{
Work at the Ames Laboratory was supported by the Department of Energy - Basic
Energy Sciences under Contract No. DE-AC02-07CH11358.}

\appendix

\section{$A_z$, $B_x$, and $B_y$ for a cylinder of rectangular cross section
carrying uniform current density}

In the following section we shall make use of several auxiliary
functions.  Consider the vector potential  $\mathbf
A(x,z) = \hat y A_y(x,z)$ and   magnetic induction 
$\mathbf B(x,z) = \hat x B_x(x,z) + \hat z
B_z(x,z) =
\nabla \times \mathbf A(x,z)$ generated by a uniform
current density $J_y$ in the region $x_1 < x < x_2$ and $-b < z < b$:
\begin{eqnarray}
A_y & = & (\mu_0 a^2 J_y / 2\pi) f_y(x_1',x_2',b',x',z'), \\
B_x & = & (\mu_0 a J_y / 2\pi) f_x(x_1',x_2',b',x',z'), \\
B_z & = & (\mu_0 a J_y / 2\pi) f_z(x_1',x_2',b',x',z'),
\end{eqnarray}
where the primes denote dimensionless variables, $x_1' = x_1/a$, etc.
Since $B_z = \partial A_y /\partial x$ and  
$B_z = -\partial A_y /\partial z$, we have 
$f_z = \partial f_y/\partial x'$ and
$f_x = -\partial f_y/\partial z'$.
The dimensionless functions $f_y$, $f_x$, and $f_z$ are (dropping the primes for
simplicity):
\begin{eqnarray}
\fl
f_y(x_1,x_2,b,x,z)=\frac{1}{2}\Big[-(x-x_1)(z+b)\ln[(x-x_1)^2+(z+b)^2] \nonumber \\
+(x-x_1)(z-b)\ln[(x-x_1)^2+(z-b)^2] \nonumber \\
+(x-x_2)(z+b)\ln[(x-x_2)^2+(z+b)^2] \nonumber \\-
(x-x_2)(z-b)\ln[(x-x_2)^2+(z-b)^2] \nonumber \\
-(x-x_1)^2\arctan\big(\frac{z+b}{x-x_1}\Big)
+(x-x_1)^2\arctan\big(\frac{z-b}{x-x_1}\Big)\nonumber \\
+(x-x_2)^2\arctan\big(\frac{z+b}{x-x_2}\Big)
-(x-x_2)^2\arctan\big(\frac{z-b}{x-x_2}\Big)\nonumber \\
-(z+b)^2\arctan\big(\frac{x-x_1}{z+b}\Big)
+(z+b)^2\arctan\big(\frac{x-x_2}{z+b}\Big)\nonumber \\
+(z-b)^2\arctan\big(\frac{x-x_1}{z-b}\Big)
-(z-b)^2\arctan\big(\frac{x-x_2}{z-b}\Big)\Big],
\label{fy} \\
\fl
f_x(x_1,x_2,b,x,z)=(x-x_1)\ln\sqrt{(x-x_1)^2+(z+b)^2}-
(x-x_1)\ln\sqrt{(x-x_1)^2+(z-b)^2} \nonumber \\
-(x-x_2)\ln\sqrt{(x-x_2)^2+(z+b)^2}+
(x-x_2)\ln\sqrt{(x-x_2)^2+(z-b)^2} \nonumber \\
+(z+b)\arctan\big(\frac{x-x_1}{z+b}\Big)
-(z+b)\arctan\big(\frac{x-x_2}{z+b}\Big)\nonumber \\
-(z-b)\arctan\big(\frac{x-x_1}{z-b}\Big)
+(z-b)\arctan\big(\frac{x-x_2}{z-b}\Big),\\
\fl
f_z(x_1,x_2,b,x,z)=(z+b)\ln\sqrt{(x-x_1)^2+(z+b)^2}-
(z+b)\ln\sqrt{(x-x_2)^2+(z+b)^2} \nonumber \\
-(z-b)\ln\sqrt{(x-x_1)^2+(z-b)^2}+
(z-b)\ln\sqrt{(x-x_2)^2+(z-b)^2} \nonumber \\
+(x-x_1)\arctan\big(\frac{z+b}{x-x_1}\Big)
-(x-x_1)\arctan\big(\frac{z-b}{x-x_1}\Big)\nonumber \\
-(x-x_2)\arctan\big(\frac{z+b}{x-x_2}\Big)
+(x-x_2)\arctan\big(\frac{z-b}{x-x_2}\Big).
\end{eqnarray}

Although these functions appear to have singularities in the $\ln$ and
$\arctan$ terms whenever the point $(x,z)$ is on one of the boundaries ($x =
x_1$,
$x = x_2$, $z = -b$, or $z = b$), the prefactors [$(x-x_1)$, $(x-x_2)$,
$(z+b)$, or $(z-b)$] cause these terms to vanish there.  Corresponding to the
conditions that $\nabla \cdot \mathbf B = 0$ and $\mathbf J = \nabla \times
\mathbf B /\mu_0$, we have $\partial f_x/\partial x +\partial f_z/\partial z =
0$ and 
\begin{eqnarray}
\frac{\partial f_x}{\partial z} -\frac{\partial f_z}{\partial x} &=& 
2 \pi, \;x_1 <
x < x_2
\; {\rm and} \;|z| < b,\\ &=& 0, \;  {\rm otherwise.}
\end{eqnarray}

\section{Finite $Z$ stack for constant $c$}

We next wish to calculate the vector potential  $\mathbf
A(x,z) = \hat y A_y(x,z)$ and the corresponding magnetic induction 
$\mathbf B(x,z) = \hat x B_x(x,z) + \hat z
B_z(x,z) =
\nabla \times \mathbf A(x,z)$
generated by the following current densities $J_y$ in a stack of height $2b$: 
$J_y = J_c$ for $c < |x| < a$ and 
$J_y= J_m= j_m J_c$  for
$|x|<c$, where
\begin{equation}
j_m = [1-(a/c)(1-I/I_c)].
\label{jm}
\end{equation}
The fields can be expressed as sums of contributions from the three
regions,  $-a < x < -c$,   $-c < x < c$, and $c < x < a$, each of total height
$2b$, where $-b < z < b$:
\begin{eqnarray}
\fl
A_y(x,z)  =  (\mu_0 a^2 J_c / 2\pi) [f_y(-1,-c',b',x',z')
+j_m f_y(-c',c',b',x',z') + f_y(c',1,b',x',z')], \nonumber \\
\label{Ay}\\
\fl
B_x(x,z)  =  (\mu_0 a J_c / 2\pi)[f_x(-1,-c',b',x',z')
+j_m f_x(-c',c',b',x',z') + f_x(c',1,b',x',z')], \nonumber \\
\\
\fl
B_z(x,z)  =  (\mu_0 a J_c / 2\pi) [f_z(-1,-c',b',x',z')
+j_m f_z(-c',c',b',x',z') + f_z(c',1,b',x',z')],\nonumber \\
\label{B4}
\end{eqnarray}
where $x' = x/a$, $z' = z/a,$ $b' = b/a$ and $c' = c/a$.
For  given values of $b' = b/a$ and $I/I_c$, we determine $c$ using criterion (i)
by requiring that 
\begin{equation}
\int_0^b dz \int_0^c dx B_z(x,z) = \int_0^b dz [A_y(c,z)-A_y(0,z)] = 0,
\label{Aintegral}
\end{equation}
which  can be solved numerically, using analytic expressions for
the integrals of
$f_y$ required in  (\ref{Aintegral}). Figure
\ref{Fig03} shows plots of
$c' = c/a$, obtained from (\ref{jm}), (\ref{Ay}), and  (\ref{Aintegral}),
vs $I/I_c$ for various values of $b' = b/a$, and figure  \ref{Fig04} shows
corresponding plots of
$j_m$.   

Contours of constant
$A_y(x,z)$ obtained from  (\ref{Ay}) correspond to magnetic field
lines.

\section{Losses for $I/I_c \to 1$}

In the limit
as  $i = I/I_c
\to 1$ [see figure 3], $c \to 0$ independent of either the value of $b/a$ or
the criterion (i)-(v) used to determine $c$. 
We can then apply the anisotropic homogeneous-medium approximation to calculate $Q' =
4 Q_{init}'$ from (2) and (\ref{B4}) using $c' = 0$ and
$f_z(0,0,b',x',z') = 0$.  The result for $I/I_c = 1$ is
\begin{equation}
Q' = \frac{8}{\pi}\mu_0J_c^2a^4 f(b/a),
\label{C1}
\end{equation}
where
\begin{eqnarray}
f(u)& =& \frac{1}{3}\Big[-3u^2 + 8u(1-u^2)\tan^{-1}(u) -2u(3-4u^2)\tan^{-1}(2u)
\nonumber \\
&&+u^4\ln\Big(\frac{u^2}{1+u^2}\Big) 
+ 6u^2
\ln\Big(\frac{4+4u^2}{1+4u^2}\Big)  +\ln\Big(\frac{\sqrt{1+4u^2}}{1+u^2}\Big)\Big],
\end{eqnarray}
Expansions of $f(u)$ about $u = \infty$ and $u = 0$ yield the leading terms,
\begin{eqnarray}
f(u)&=& \pi u/3, \;  \;\;u \to \infty, 
\label{C3}\\
 &=& 2(\ln 4 - 1) u^2, \; 
\;\;u \to 0.
\label{C4}
\end{eqnarray}

When $I/I_c = 1$, we obtain from  (\ref{C1}) and (3) the ratio 
\begin{equation}
R_1 =\frac{Q'}{Q_{inf}'} = \frac{3f(b/a)}{\pi(b/a)},
\label{R1}
\end{equation}
which is plotted vs $b/a$ as the upper curve in figure 15.  When $I/I_c = 1$
and $b/a \to \infty$, we find from  (\ref{C3}) and (\ref{R1}) that $Q'/Q_{inf}'
\to 1$, as expected.  When $I/I_c = 1$
and $b/a \to 0$, we find from  (\ref{C1}), (\ref{C4}), and (\ref{QNorris}) that
$Q'/Q_{strip}'
\to 1$.

\section{Losses for $I/I_c \to 0$}

According to the above anisotropic homogeneous-medium approximation with $c$
independent of $z$, the ac loss per cycle per unit length $Q' =4 Q_{init}'$ 
(\ref{Qinit}) is proportional to $(I/I_c)^3$ in the limit
as  $i = I/I_c
\to 0.$  In this limit we have $c' =1-\epsilon$
and [from  (\ref{jm})] $j_m = i-\epsilon$ to  first order in
$\epsilon$, where $\epsilon = k i$ and $k$ is a constant of order unity ($k < 1$),
which depends upon the criterion used to determine $c$.  By expanding $f_y$ in
 (\ref{Ay}) or
$f_z$ in  (\ref{B4}) to first order in $\epsilon$ and carrying out the integration in
 (\ref{Aintegral}), we obtain for the value of  $k$ using criterion (i) 
\begin{eqnarray}
k_i &=& [8\pi
b'^3\!-48b'^2\ln2-16b'(3\!-\!b'^2)\!\tan^{-1}b'
+8b'(3\!-\!4b'^2)\!\tan^{-1}(2b') \nonumber \\
&&
+8(1\!-\!3b'^2)\!\ln(1\!+\!b'^2)-2(1\!-\!12b'^2)\!\ln(1\!+\!4b'^2)]/
 \nonumber \\
&&[8\pi b'^3
+16b'^3\!\tan^{-1}b'\!\! 
-8b'(3\!+\!4b'^2)\!\tan^{-1}(2b')+12b'^2\!\ln(b'^2)
\nonumber \\
&&
-4(1\!+\!3b'^2)\!\ln(1+b'^2)+4\ln(1\!+\!4b'^2)].
\label{ki}
\end{eqnarray}
Similarly, by expanding
$f_z$ in  (\ref{B4}) to first order in $\epsilon$ and setting $B_z(c,0)=0$, we obtain
for criterion (ii) 
\begin{equation}
k_{ii} =
\frac{4\tan^{-1}(b'/2)+b'\ln(1\!+\!4/b'^2)}{\pi+2 \tan^{-1}(b'/2)+b'\ln(1\!+\!4/b'^2)}.
\label{kii}
\end{equation}
Carrying out the integration required in  (\ref{Qinit}), we obtain in the limit
as  $i = I/I_c
\to 0$,
\begin{equation}
Q' = \frac{4}{\pi}\mu_0 J_c^2a^4g(k,b')(I/I_c)^3,
\label{Q'smalli}
\end{equation}
where
\begin{eqnarray}
g(k,b')&=&[4b'\tan^{-1}b'+b'^2\ln(1+1/b'^2)-\ln(1+b'^2)]k^2 \nonumber \\
&&-[\pi b'/3 +2 b' \tan^{-1}b'+b'^2\ln(1+1/b'^2)]k^3.
\end{eqnarray}
The ratio of the result in  (\ref{Q'smalli}) to $Q'_{inf}$ (\ref{Qinf}) is 
\begin{equation}
R_0 = \frac{Q'}{Q_{inf}'} = \frac{3}{2\pi}\frac{g(k,b')}{b'},
\label{R0}
\end{equation}
which is plotted as the lower curves in figure 15 for two of the criteria [(i) solid
and (ii) dashed] used to determine the constant $c$.  As $b' \to \infty$, $k_i \to
1$, 
$k_{ii} \to 1,$ and $R_0 \to 1$ for both criteria, and as $b' \to 0$, $k_i \to
0$, 
$k_{ii} \to 0,$ and $R_0 \to 0$ for both criteria.  However, for very small values of
$b'$, where the present approach is not accurate, the value of $R_0$
for $k = k_i$ is negative for $0 < b' < 0.0457$, an unphysical result.

\section{Contribution to the losses from the middle region, $|x| < c$}

In the above sections we have calculated the ac losses in the outer regions ($c < |x| <
a$) due to perpendicular magnetic flux ($\propto B_z$) moving in and out
from the edges of the tapes.  We have so far neglected the ac losses in the middle
region ($|x| < c$) due to parallel magnetic flux ($\propto B_x$) moving in and out from
the top and bottom surfaces of the tapes, on the assumption that these losses are very
small.  In this appendix we present equations that can be used to confirm that the
losses in the middle region are indeed much smaller than those in the outer
regions.  

Consider the finite $Z$ stack sketched in figure 1, and label the $z$ coordinate of
a given tape as $z_n$.  For an odd number of tapes in the stack $z_n = nD$, $n =
0,\; 
\pm 1,\; \pm 2, ...,$ and for an even number of tapes $z_n = \pm D/2, \; \pm3 D/2, \;
\pm 5D/2, ...\;  .$  In the middle region
$|x| < c$, each tape carries an ac current of reduced amplitude
$j_m = J_m/J_c$, normalized to the critical current density.  However, since $\partial
B_x/\partial z =
\mu_0 J_m,$ we have $B_x = \mu_0 J_m z$, such that the tape at $z_n$ is also subject
to an applied in-phase ac magnetic induction of amplitude $\mu_0 J_m |z_n|$.  It is
convenient to normalize this to the penetration field $B_p = \mu_0 j_c d/2 = \mu_0 J_c
D/2$, such that the applied field has the reduced amplitude $h_n = j_m |z_n|/(D/2)$.  

$Q_v$, the hysteretic loss per cycle per unit volume of a superconducting slab
subjected to an ac parallel field of reduced amplitude $h$ and an in-phase current of
reduced amplitude $i =
\bar j/j_c < 1$, has been calculated by Carr \cite{Carr79}, whose results can be
expressed as 
\begin{equation}
Q_v = \frac{2 B_p^2}{3\mu_0}f(h,i),
\end{equation}
where 
\begin{eqnarray}
f(h,i)&=&i^2(i+3h), \;h \le i, 
\label{E2} \\
&=&h(h^2+3i^2), \; i \le h < 1, 
\label{E3} \\
&=&h(3+i^2)-2(1-i^3)+\frac{6i^2(1-i)^2}{(h-i)}-\frac{4i^2(1-i)^3}{(h-i)^2}, \; h \ge 1.
\label{E4} 
\end{eqnarray}

\begin{figure}
\centering
\includegraphics[width=8cm]{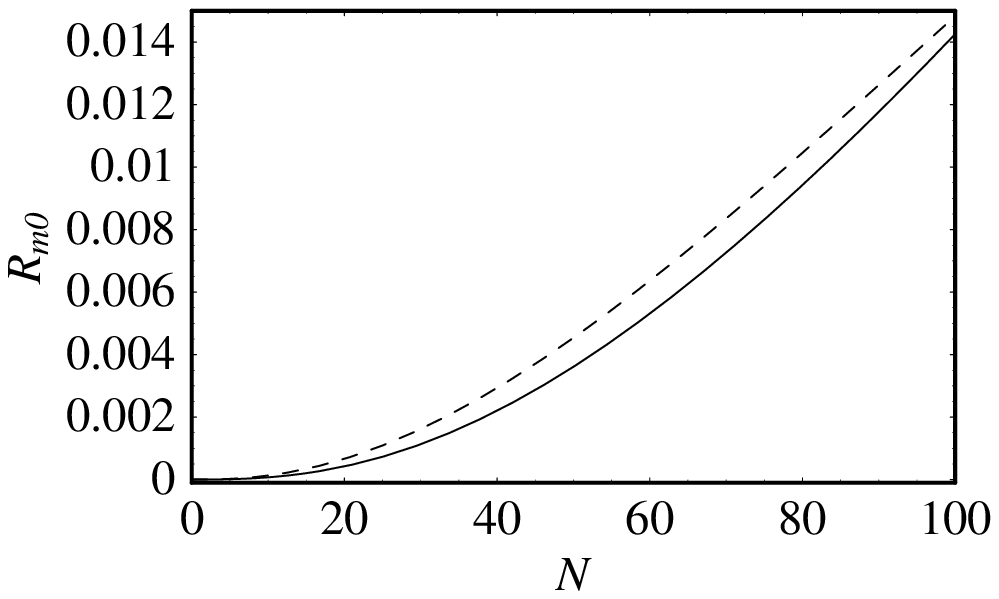}
\caption{$R_{m0}$ (E.8), the contribution $Q_m'$ (E.6) to the ac loss per
cycle per unit length due to the middle region ($|x| < c$) of the
$Z$ stack, normalized to
$Q_{inf}'$ (\ref{Qinf}), the ac loss per cycle per unit length for an
equivalent cross section  of an
infinite slab, in the limit as $I/I_c \to 0$, vs $N$, the number of tapes in the stack,
for  $d' = 2 \times 10^{-4}$, $D' = 2 \times 10^{-2}$, and $N = 100 b' = 100 b/a$.  
The results were calculated using two different criteria [(i) solid  and (ii) dashed]
for calculating the parameter $c$.}
\label{Rm0Plot}
\end{figure}

\begin{figure}
\centering
\includegraphics[width=8cm]{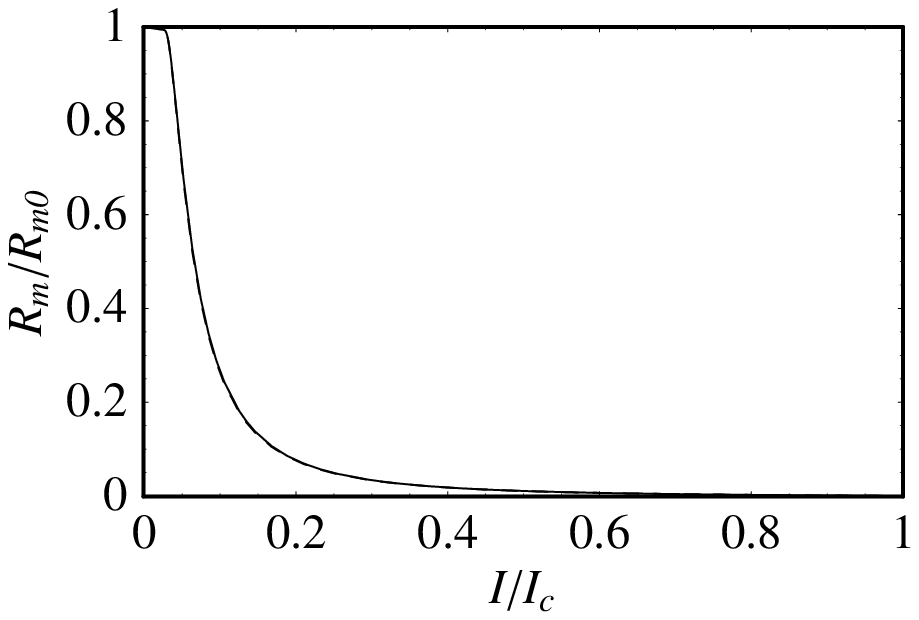}
\caption{The ratio $R_{m}/R_{m0}$, calculated from  (E.7) and (E.8), vs $I/I_c$
assuming $d' = 2 \times 10^{-4}$, $D' = 2 \times 10^{-2}$, and $N = 100$ or $b' = 
b/a = 1$.  Two different criteria [(i)   and (ii)]
for calculating the parameter $c$ yield nearly identical results.}
\label{RmratioPlot}
\end{figure}

$q_n'$, the hysteretic loss in tape $n$ per cycle per unit length  due to the ac
current and parallel field in the middle region ($|x| < c$) of the stack, can
therefore be calculated from 
\begin{equation}
q_n' = Q_v 2cd = \frac{4 B_p^2 c d}{3\mu_0}f(h_n,j_m),
\end{equation}
and $Q_m'$, the total hysteretic loss per cycle per unit length of the $Z$ stack due
to  the ac
currents and parallel fields in the middle region ($|x| < c$), can be obtained by
carrying out the sum over all tapes, 
\begin{equation}
Q_m' = \sum_n q_n'.
\end{equation}
As can be seen from 
(\ref{E2})-(\ref{E4}),  the appropriate expression for $f(h_n,j_m)$ to be used
in the sum depends upon the value of $h_n$ relative to $j_m$ and 1.
The ratio of $Q_m'$ to $Q'_{inf}$ (\ref{Qinf}) is 
\begin{equation}
R_m = \frac{Q'_m}{Q'_{inf}} = \frac{cdD^2}{8a^3b(I/I_c)^3}\sum_n f(h_n,j_m).
\label{Rm}
\end{equation}
Numerical evaluation shows that, in contrast to the behavior of $R_0$ and $R_1$ shown
in figure 15,
$R_m$ is a monotonically {\it decreasing} function of
$I/I_c$ with its  maximum value at $I/I_c = 0$,
given (for an even number
$N$ of tapes) by 
\begin{equation}
R_{m0}=d' D' N(N^2 + 4) (1-k)^3/16,
\end{equation}
where $d' = d/a$, $D' = D/a$, and $k$ [see  (\ref{ki}) and
(\ref{kii})] depends upon the criterion used to determine
$c$. To estimate the order of magnitude of the middle-region losses we use the
following assumptions: $2a = 10$ mm, $D = 100\; \mu$m, $d  = 1\; \mu$m, $2b = N D$,
such that $d' = 2 \times 10^{-4}$, $D' = 2 \times 10^{-2}$, and $N = 100 b' = 100
b/a$.  Shown in figure \ref{Rm0Plot} is a plot of $R_{m0}$ vs $N$, the number of
tapes in the stack, for the two criteria [(i) solid and (ii) dashed] we have used to
calculate
$c$.  Figure \ref{RmratioPlot}, calculated for $b' = 1$, shows the general behavior
of how the ratio
$R_{m}/R_{m0}$ depends upon
$I/I_c$.  This ratio has its maximum as $I/I_c \to 0$, where $c' = c/a \to 1$, but 
vanishes as  $I/I_c \to 1$, where $c' = c/a \to 0.$  Although both criteria (i) and
(ii) for choosing
$c$ were used to calculate this ratio, the two curves
are indistinguishable on this plot. 
Comparing figures \ref{Rm0Plot} and \ref{RmratioPlot} for the middle-region losses
with figures 14 and 15 for the outer-region losses, we see that for $I/I_c > 0.2$ the
hysteretic losses from the middle region of the stack
$|x| < c$ are typically at least two orders of magnitude smaller than those from the
edges of the tapes.  Howeover, in the limit as $I/I_c \to 0,$ where the middle region
includes nearly the entire volume and the outer regions shrink to zero, the
middle-region losses become more important but still remain relatively small.  In
summary, these results confirm that for $I/I_c > 0.2$ the hysteretic losses from the
middle region of the stack
$|x| < c$ are typically several orders of magnitude smaller than those from the edges
of the tapes.  

\section*{References}
\begin{thebibliography}{99}
\bibitem{Polak06} Polak M, Demencik E, Jansak L, Mozola P, Aized D,
Thieme C L H, Levin G A and Barnes PN 2006 {\it Appl. Phys.
Lett.} {\bf 88} 232501 
\bibitem{Grilli07} Grilli F and Ashworth SP 2007 {\it Supercond. Sci.
Technol.} {\bf 20} 794 
\bibitem{Claassen06} Claassen J H 2006 {\it Appl. Phys. Lett.} {\bf 88} 122512
\bibitem{Pardo05} Pardo E, Sanchez A, Chen D-X and Navau C 2005
\PR B {\bf 71} 134517
\bibitem{Goyal05} 2005 {\it Second-Generation HTS Conductors} ed A Goyal
(Boston: Kluwer)
\bibitem{Mawatari97} 
Mawatari Y 1997 in {\it Advances in Superconductivity IX}
ed S Nakajima S and M Murakami (Tokyo: Springer) p 575
\bibitem{Muller97} 
M\"uller K-H 1997 {\it Physica} C {\bf289} 123
\bibitem{Muller99}
M\"uller K-H 1999 {\it Physica} C {\bf312} 149 
\bibitem{Bean62} 
Bean C P 1962 {\it Phys. Rev. Lett.} {\bf 8} 250 
\bibitem{Bean64} Bean C P 1964 {\it Rev. Mod.
Phys.} {\bf 36} 31
\bibitem{Brandt93b}
Brandt 	E H and Indenbom M 1993 \PR B {\bf 48} 12 893
\bibitem{Zeldov94b}
Zeldov E, Clem J R, McElfresh M and Darwin M 1994 \PR B
{\bf 49} 9802 
\bibitem{Mawatari06}  Mawatari Y and Kajikawa K 2006 {\it Appl. Phys. Lett.}
{\bf 88} 092503  
\bibitem{Mawatari07} Mawatari Y and Kajikawa K 2007 {\it Appl. Phys. Lett.}
{\bf 90} 022506 
\bibitem{Norris70} Norris W T 1970 \JPD {\bf 3} 489 
\bibitem{Halse70} Halse M R 1970 \JPD {\bf 3} 717 
\bibitem{Math} 2005 {\it Mathematica, Version 5.2} (Champaign, IL: Wolfram
Research)
\bibitem{Carr79} Carr, Jr., W J 1979 {\it IEEE Trans. Magn.} {\bf MAG-15} 240
\end{thebibliography}
\end{document}